\pdfoutput=1
\documentclass[11pt]{article}

\usepackage[preprint]{acl}

\usepackage{times}
\usepackage{latexsym}
\usepackage{multicol}
\usepackage{hyperref}
\usepackage[T1]{fontenc}
\usepackage[utf8]{inputenc}

\usepackage{microtype}

\usepackage{inconsolata}

\usepackage{graphicx}
\usepackage{amsfonts}
\usepackage{amsmath}
\usepackage{amssymb}
\usepackage{booktabs}
\usepackage[capitalize,noabbrev,nameinlink]{cleveref}

\title{\texttt{BM25S}: Orders of magnitude faster lexical search via eager sparse scoring}

\author{Xing Han Lù \\
  McGill University and Mila Quebec AI Institute \\
  \texttt{xing.han.lu@mail.mcgill.ca}}

\begin{document}
\maketitle
\begin{abstract}
We introduce \texttt{BM25S}, an efficient Python-based implementation of BM25 that only depends on Numpy\footnote{\url{https://numpy.org/}} and Scipy\footnote{\url{https://scipy.org/}}. \texttt{BM25S} achieves up to a 500x speedup compared to the most popular Python-based framework by eagerly computing BM25 scores during indexing and storing them into sparse matrices.
It also achieves considerable speedups compared to highly optimized Java-based implementations, which are used by popular commercial products.
Finally, \texttt{BM25S} reproduces the exact implementation of five BM25 variants based on \citet{kamphuis2020_which_bm25_do_you_mean} by extending eager scoring to non-sparse variants using a novel score shifting method.
The code can be found at \url{https://github.com/xhluca/bm25s}
\end{abstract}

\section{Background}

Sparse lexical search algorithms, such as the BM25 family \cite{robertson1995okapi} remain widely used as they do not need to be trained, can be applied to multiple languages, and are generally faster, especially when using highly efficient Java-based implementations. Those Java implementations, usually based on Lucene\footnote{\url{https://lucene.apache.org/}}, are accessible inside Python via the Pyserini reproducibility toolkit \cite{Lin2021PyseriniAP}, and through HTTP by using the Elasticsearch web client\footnote{\url{https://www.elastic.co/elasticsearch}}. The Lucene-based libraries are known to be faster than existing Python-based implementations, such as Rank-BM25\footnote{\url{https://github.com/dorianbrown/rank_bm25}}.

This work shows that it is possible to achieve a significant speedup compared to existing Python-based implementations by introducing two improvements: eagerly calculating all possible scores that can be assigned to any future query token when indexing a corpus, and storing those calculations inside sparse matrices to enable faster slicing and summations. The idea of sparse matrices was previously explored in BM25-PT\footnote{\url{https://github.com/jxmorris12/bm25_pt}}, which pre-computes BM25 scores using PyTorch and multiplies them with a bag-of-word encoding of the query via sparse matrix multiplication.

This work expands upon the initial idea proposed by the BM25-PT project by significantly simplifying the implementation and introducing a strategy to generalize to other variants of the original BM25. Unlike BM25-pt, \texttt{BM25S} does not rely on PyTorch, and instead uses Scipy's sparse matrix implementation. Whereas BM25-PT multiplies bag-of-words with the document matrix, \texttt{BM25S} instead slices relevant indices and sums across the token-dimension, removing the need of matrix multiplications.

At the implementation level, \texttt{BM25S} also introduces a simple but fast Python-based tokenizer that combines Scikit-Learn's text splitting \cite{Pedregosa2011ScikitlearnML}, Elastic's stopword list\footnote{\url{https://www.elastic.co/guide/en/elasticsearch/guide/current/stopwords.html}}, and (optionally) integrates a C-based implementation of the Snowball stemmer \cite{2014SnowballSB}. This achieves a better performance compared to subword tokenizers \cite{Kudo2018SentencePieceAS} used by BM25-PT. Finally, it implements top-k retrieval using an average $O(n)$ time complexity when selecting the $K$ most relevant documents from a set of $n$ scores associated with each document.

\section{Implementation}

The implementation described below follows the study by \citet{kamphuis2020_which_bm25_do_you_mean}.

\paragraph{Calculation of BM25} Many variants of BM25 exist, which could lead to significant confusion about the exact scoring method used in a given implementation \cite{kamphuis2020_which_bm25_do_you_mean}. By default, we use the scoring method proposed by Lucene. Thus, for a given query $Q$ (tokenized into $q_1,\ldots,q_{|Q|}$) and document $D$ from collection $C$, we compute the following score\footnote{We follow notations by \citet{kamphuis2020_which_bm25_do_you_mean}}:
\begin{align*}
    B(Q, D) 
    &= \sum_{i=1}^{|Q|} S(q_i, D) \\
    &= \sum_{i=1}^{|Q|} \text{IDF}(q_i, C) \frac{\text{TF}(q_i, D)}{\mathcal{D}}
\end{align*}
where $\mathcal{D} = \text{TF}(t, D) + k_1 \left(1 - b + b \frac{|D|}{L_{avg}} \right)$, $L_{avg}$ is the average length of documents in corpus $C$ (calculated in number of tokens), $\text{TF}(q_i, D)$ is the term frequency of token $q_i$ within the set of tokens in $D$. The \textit{IDF} is the inverse document frequency, which is calculated as:
\begin{equation*}
    \text{IDF}(q_i, C) = \ln \left(\frac { |C| - \text{DF}(q_i, C) + 0.5} { \text{DF}(q_i, C) + 0.5 } + 1 \right)
\end{equation*}
Where document frequency $\text{DF}(q_i, C)$ is the number of documents in $C$ containing $q_i$. 
Although $B(Q,D)$ depends on the query, which is only given during retrieval, we show below how to reformulate the equation to eagerly calculate the TF and IDF during indexing.

\paragraph{Eager index-time scoring}
Let's now consider all tokens in a vocabulary $V$, denoted by $t \in V$. We can reformulate $S(t, D)$ as:
\begin{equation*}
    S(t, D) = \text{TF}(t, D) \cdot \text{IDF}(t, C) \frac{1}{\mathcal{D}}
\end{equation*}
When $t$ is a token that is not present in document $D$, then $\text{TF}(t, D)=0$, leading to $S(t, D) = 0$ as well. This means that, for most tokens in vocabulary $V$, we can simply set the relevance score to 0, and only compute values for $t$ that are actually in the document $D$. This calculation can be done during the indexing process, thus avoiding the need to compute $S(q_i, D)$ at query time, apart from straightforward summations.

\paragraph{Assigning Query Scores}
Given our sparse matrix of shape $|V| \times |C|$, we can use the query tokens to select relevant rows, leaving us a matrix of shape $|Q| \times |C|$, which we can then sum across the column dimension, resulting in a single $|C|$-dimension vector (representing the score of the score of each document for the query).

\paragraph{Efficient Matrix Sparsity} We implement a sparse matrix in \textit{Compressed Sparse Column} (CSC) format (\texttt{scipy.sparse.csc\_matrix})\footnote{\url{https://docs.scipy.org/doc/scipy/reference/generated/scipy.sparse.csc_matrix.html}}, which provides an efficient conversion between the coordinate and CSC format. Since we slice and sum alongside the column dimension, this implementation is the optimal choice among sparse matrix implementations. In practice, we replicate the sparse operations directly using Numpy array.

\paragraph{Tokenization} To split the text, we use the same Regular Expression pattern used by Scikit-Learn \cite{Pedregosa2011ScikitlearnML} for their own tokenizers, which is \verb|r"(?u)\b\w\w+\b"|. This pattern conveniently parses words in UTF-8 (allowing coverage of various languages), with \verb|\b| handling word boundaries. Then, if stemming is desired, we can stem all words in the vocabulary, which can be used to look up the stemmed version of each word in the collection. Finally, we build a dictionary mapping each unique (stemmed) word to an integer index, which we use to convert the tokens into their corresponding index, thus significantly reducing memory usage and allowing them to be used to slice Scipy matrices and Numpy arrays.

\paragraph{Top-k selection} Upon computing scores for all documents in a collection, we can complete the search process by selecting the top-$k$ most relevant elements. A naive approach to this would be to sort the score vector and select the last $k$ elements; instead, we take the partition of the array, selecting only the last $k$ documents (unordered). Using an algorithm such as Quickselect \cite{hoare_quickselect}, we can accomplish this in an average time complexity of $O(n)$ for $n$ documents in the collection, whereas sorting requires $O(n \log n)$. If the user wishes to receive the top-$k$ results in order, sorting the partitioned documents would take an additional $O(k \log k)$, which is a negligible increase in time complexity assuming $k \ll n$. In practice, \texttt{BM25S} allows the use of two implementations: one based in \texttt{numpy}, which leverages \texttt{np.argpartition}, and another in \texttt{jax}, which relies on XLA's top-k implementation. Numpy's \texttt{argpartition} uses\footnote{\url{https://numpy.org/doc/stable/reference/generated/numpy.argpartition.html}} the introspective selection algorithm \cite{musser1997introspective}, which modifies the quickselect algorithm to ensure that the worst-case performance remains in $O(n)$. Although this guarantees optimal time complexity, we observe that JAX's implementation achieves better performance in practice.

\paragraph{Multi-threading} We implement optional multi-threading capabilities through pooled executors\footnote{Using \texttt{concurrent.futures.ThreadPoolExecutor}} to achieve further speed-up during retrieval.

\paragraph{Alternative BM25 implementations}
Above, we describe how to implement \texttt{BM25S} for one variant of BM25 (namely, Lucene). However, we can easily extend the \texttt{BM25S} method to many variants of BM25; the sparsity can be directly applied to Robertson's original design \cite{robertson1995okapi}, ATIRE \cite{ATIRE_Trotman2014ImprovementsTB}, and Lucene. For other models, a modification of the scoring described above is needed.

\begin{table}[t]
    \centering
    \footnotesize
    \begin{tabular}{lrrrr}
    \toprule
    Dataset & BM25S & ES & PT & Rank \\
    \midrule
    ArguAna & \textbf{573.91} & 13.67 & \textbf{110.51} & 2.00 \\
    Climate-FEVER & 13.09 & 4.02 & OOM & 0.03 \\
    CQADupstack & \textbf{170.91} & 13.38 & OOM & 0.77 \\
    DBPedia & 13.44 & 10.68 & OOM & 0.11 \\
    FEVER & 20.19 & 7.45 & OOM & 0.06 \\
    FiQA & \textbf{507.03} & 16.96 & 20.52 & 4.46 \\
    HotpotQA & 20.88 & 7.11 & OOM & 0.04 \\
    MSMARCO & 12.20 & 11.88 & OOM & 0.07 \\
    NFCorpus & \textbf{1196.16} & 45.84 & \textbf{256.67} & \textbf{224.66 }\\
    NQ & 41.85 & 12.16 & OOM & 0.10 \\
    Quora & \textbf{183.53} & 21.80 & 6.49 & 1.18 \\
    SCIDOCS & \textbf{767.05} & 17.93 & 41.34 & 9.01 \\
    SciFact & \textbf{952.92} & 20.81 & \textbf{184.30} & 47.60 \\
    TREC-COVID & \textbf{85.64} & 7.34 & 3.73 & 1.48 \\
    Touche-2020 & \textbf{60.59} & 13.53 & OOM & 1.10 \\
    \bottomrule
    \end{tabular}
    \caption{\footnotesize{To calculate the throughput, we calculate the number of \textit{queries per second (QPS)} that each model can process for each task in the public section of the BEIR leaderboard; instances achieve over 50 QPS are shown in \textbf{bold}. We compare \texttt{BM25S}, BM25-PT (PT), Elasticsearch (ES) and Rank-BM25 (Rank). OOM indicates failure due to out-of-memory issues.}}
    \label{tab:qps}
\end{table}

\begin{table*}[t]
    \footnotesize
    \setlength{\tabcolsep}{4.5pt}
    \centering
\begin{tabular}{llccccccccccccccccc}
\toprule
Stop & Stem & Avg. & AG & CD & CF & DB & FQ & FV & HP & MS & NF & NQ & QR & SD & SF & TC & WT \\
\midrule
Eng. & None & 38.4 & 48.3 & 29.4 & 13.1 & 27.0 & 23.3 & 48.2 & 56.3 & 21.2 & 30.6 & 27.3 & 74.8 & 15.4 & 66.2 & 59.5 & 35.8 \\
Eng. & Snow. & 39.7 & 49.3 & 29.9 & 13.6 & 29.9 & 25.1 & 48.1 & 56.9 & 21.9 & 32.1 & 28.5 & 80.4 & 15.8 & 68.7 & 62.3 & 33.1 \\
None & None & 38.3 & 46.8 & 29.6 & 13.6 & 26.6 & 23.2 & 48.8 & 56.9 & 21.1 & 30.6 & 27.8 & 74.2 & 15.2 & 66.1 & 58.3 & 35.9 \\
None & Snow. & 39.6 & 47.7 & 30.2 & 13.9 & 29.5 & 25.1 & 48.7 & 57.5 & 21.7 & 32.0 & 29.1 & 79.7 & 15.6 & 68.5 & 61.6 & 33.4 \\
\bottomrule
\end{tabular}

    \caption{NDCG@10 results of different tokenization schemes (including and excluding stopwords and the Snowball stemmer) on all BEIR dataset (\cref{sec:appendix} provides a list of datasets). We notice that including both stopwords and stemming modestly improves the performance of the BM25 algorithm.}
    \label{tab:comparison_tokenizer_effect}
\end{table*}

\subsection{Extending sparsity via non-occurrence adjustments}

For BM25L \cite{BM25L_Lv2011AdaptiveTF}, BM25+ \cite{BM25L_Lv2011AdaptiveTF} and $\text{TF}_{l \circ \delta \circ p}\times \text{IDF}$ \cite{TFIDF_Rousseau2013CompositionOT}, we notice that when $TF(t, D)=0$, the value of $S(t,D)$ will not be zero; we denote this value as a scalar\footnote{We note that it is not an $|D|$-dimensional array since it does not depend on $D$, apart from the document frequency of $t$, which can be represented with a $|V|$-dimensional array.}
 $S^{\theta}(t)$, which represents the score of $t$ when it does not occur in document $D$.

Clearly, constructing a $|V| \times |C|$ dense matrix would use up too much memory\footnote{For example, we would need 1.6TB of RAM to store a dense matrix of 2M documents with 200K words in the vocabulary.}. Instead, we can still achieve sparsity by subtracting $S^{\theta}(t)$ from each token $t$ and document $D$ in the score matrix (since most tokens $t$ in the vocabulary will not be present in any given document $D$, their value in the score matrix will be 0).
Then, during retrieval, we can simply compute $S^{\theta}(q_i)$ for each query $q_i \in Q$, and sum it up to get a single scalar that we can add to the final score (which would not affect the rank).

More formally, for an empty document $\emptyset$, we define $S^{\theta}(t) = S(t, \emptyset)$ as the nonoccurrence score for token $t$. Then, the differential score $S^\Delta (t, D)$ is defined as:
\begin{equation*}
    S^\Delta (t, D) = S(t, D) - S^{\theta}(t)
\end{equation*}
Then, we reformulate the BM25 ($B$) score as:
\begin{align*}
    B(Q,D) 
    &= \sum_{i=1}^{|Q|} S(q_i, D) \\
    &= \sum_{i=1}^{|Q|} \left( S(q_i, D) - S^{\theta}(q_i) + S^{\theta}(q_i) \right) \\
    &= \sum_{i=1}^{|Q|} \left( S^\Delta(q_i, D) + S^{\theta}(q_i) \right) \\
    &= \sum_{i=1}^{|Q|} S^\Delta(q_i, D) + \sum_{i=1}^{|Q|} S^{\theta}(q_i)
\end{align*}
where $\sum_{i=1}^{|Q|} S^\Delta(q_i, D)$ can be efficiently computed using the differential sparse score matrix (the same way as ATIRE, Lucene and Robertson) in \texttt{scipy}. Also, $\sum_{i=1}^{|Q|} S^{\theta}(q_i)$ only needs to be computed once for the query $Q$, and can be subsequently applied to every retrieved document to obtain the exact scores.

\begin{table*}[t]
    \footnotesize
    \setlength{\tabcolsep}{3.5pt}
    \centering
\begin{tabular}{lllccccccccccccccccc}
\toprule
$k_1$ & $b$ & Variant & Avg. & AG & CD & CF & DB & FQ & FV & HP & MS & NF & NQ & QR & SD & SF & TC & WT \\
\midrule
1.5 & 0.75 & BM25PT & -- & 44.9 & -- & -- & -- & 22.5 & -- & -- & -- & 31.9 & -- & 75.1 & 14.7 & 67.8 & 58.0 & -- \\
1.5 & 0.75 & PSRN & 40.0\textsuperscript{*} & 48.4 & -- & 14.2 & 30.0 & 25.3 & 50.0 & 57.6 & 22.1 & 32.6 & 28.6 & 80.6 & 15.6 & 68.8 & 63.4 & 33.5 \\
1.5 & 0.75 & R-BM25 & 39.6 & 49.5 & 29.6 & 13.6 & 29.9 & 25.3 & 49.3 & 58.1 & 21.1 & 32.1 & 28.5 & 80.3 & 15.8 & 68.5 & 60.1 & 32.9 \\
1.5 & 0.75 & Elastic & 42.0 & 47.7 & 29.8 & 17.8 & 31.1 & 25.3 & 62.0 & 58.6 & 22.1 & 34.4 & 31.6 & 80.6 & 16.3 & 69.0 & 68.0 & 35.4 \\
\midrule
1.5 & 0.75 & Lucene & 39.7 & 49.3 & 29.9 & 13.6 & 29.9 & 25.1 & 48.1 & 56.9 & 21.9 & 32.1 & 28.5 & 80.4 & 15.8 & 68.7 & 62.3 & 33.1 \\
0.9 & 0.4 & Lucene & 41.1 & 40.8 & 28.2 & 16.2 & 31.9 & 23.8 & 63.8 & 62.9 & 22.8 & 31.8 & 30.5 & 78.7 & 15.0 & 67.6 & 58.9 & 44.2 \\
1.2 & 0.75 & Lucene & 39.9 & 48.7 & 30.1 & 13.7 & 30.3 & 25.3 & 50.3 & 58.5 & 22.6 & 31.8 & 29.1 & 80.5 & 15.6 & 68.0 & 61.0 & 33.2 \\
\midrule
1.2 & 0.75 & ATIRE & 39.9 & 48.7 & 30.1 & 13.7 & 30.3 & 25.3 & 50.3 & 58.5 & 22.6 & 31.8 & 29.1 & 80.5 & 15.6 & 68.1 & 61.0 & 33.2 \\
1.2 & 0.75 & BM25+ & 39.9 & 48.7 & 30.1 & 13.7 & 30.3 & 25.3 & 50.3 & 58.5 & 22.6 & 31.8 & 29.1 & 80.5 & 15.6 & 68.1 & 61.0 & 33.2 \\
1.2 & 0.75 & BM25L & 39.5 & 49.6 & 29.8 & 13.5 & 29.4 & 25.0 & 46.6 & 55.9 & 21.4 & 32.2 & 28.1 & 80.3 & 15.8 & 68.7 & 62.9 & 33.0 \\
1.2 & 0.75 & Robertson & 39.9 & 49.2 & 29.9 & 13.7 & 30.3 & 25.4 & 50.3 & 58.5 & 22.6 & 31.9 & 29.2 & 80.4 & 15.5 & 68.3 & 59.0 & 33.8 \\
\bottomrule
\end{tabular}

    \caption{Comparison of different variants and parameters on all BEIR dataset (\cref{sec:appendix} provides a list of datasets). Following the recommended range of $k_1 \in [1.2, 2]$ by \citet{schutze2008introduction}, we try both $k_1=1.5$ and $k_1=1.2$ with $b=0.75$. Additionally, we use $k_1=0.9$ and $b=0.4$ following the parameters recommend in BEIR. We additionally benchmark five of the BM25 variants described in \citet{kamphuis2020_which_bm25_do_you_mean}. *note that Pyserini's average results are estimated, as the experiments for CQADupStack (CD) did not terminate due to OOM errors.}
    \label{tab:comparison_variants_bm25s}
\end{table*}

\section{Benchmarks}

\paragraph{Throughput} For benchmarking, we use the publicly available datasets from the BEIR benchmark \cite{thakur2021beir}. Results in \cref{tab:qps} show that \texttt{BM25S} is substantially faster than Rank-BM25, as it achieves over 100x higher throughput in 10 out of the 14 datasets; in one instance, it achieves a 500x speedup. Further details can be found in \cref{sec:appendix}.

\paragraph{Impact of Tokenization}
We further examine the impact of tokenization on each model in \cref{tab:comparison_tokenizer_effect} by comparing \texttt{BM25S} Lucene with $k_1=1.5$ and $b=0.75$ (1) without stemming, (2) without stop words, and (3) with neither, and (4) with both. On average, adding a Stemmer improves the score on average, wheareas the stopwords have minimal impact. However, on individual cases, the stopwords can have a bigger impact, such as in the case of Trec-COVID (TC) and ArguAna (AG). 

\paragraph{Comparing model variants}
In \cref{tab:comparison_variants_bm25s}, we compare many implementation variants, including commercial (Elasticsearch) offerings and reproducibility toolkits (Pyserini). We notice that most implementations achieve an average be between 39.7 and 40, with the exception of Elastic which achieves a marginally higher score. The variance can be attributed to the difference in the tokenization scheme; notably, the subword tokenizer used in BM25-PT likely lead to the difference in the results, considering the implementation is a hybrid between ATIRE and Lucene, both of which achieve better results with a word-level tokenizer. Moreover, although Elasticsearch is built on top of Lucene, it remains an independent commercial product, and the documentations\footnote{\url{https://www.elastic.co/guide/en/elasticsearch/reference/current/index.html}} do not clearly describe how they are splitting the text\footnote{\url{https://www.elastic.co/guide/en/elasticsearch/reference/current/split-processor.html}}, and whether they incorporate additional processing beyond the access to a Snowball stemmer and the removal of stopwords.

\section{Conclusion}

We provide a novel method for calculating BM25 scores, \texttt{BM25S}, which also offers fast tokenization out-of-the-box and efficient top-k selection during querying, minimizes dependencies and makes it usable directly inside Python. As a result, \texttt{BM25S} naturally complements previous implementations: BM25-pt can be used with PyTorch, Rank-BM25 allows changing parameters $k1$ during inference, and Pyserini provides a large collection of both sparse and dense retrieval algorithm, making it the best framework for reproducible retrieval research. On the other hand, \texttt{BM25S} remains focused on sparse and mathematically accurate implementations of BM25 that leverage the eager sparse scoring methods, with optional Python dependencies like PyStemmer for stemming and Jax for top-k selection. By minimizing dependencies, \texttt{BM25S} becomes a good choice in scenarios where storage might be limited (e.g. for edge deployments) and can be used in the browser via WebAssembly frameworks like Pyodide\footnote{\url{https://pyodide.org}} and Pyscript\footnote{\url{https://pyscript.net/}}. We believe our fast and accurate implementation will make lexical search more accessible to a broader audience.

\section*{Limitations}
A customized Python-based tokenizer (also known as analyzer) was created for \texttt{BM25S}, which allows the use of stemmer and stopwords. By focusing on a readable, extensible and fast implementation, it may not achieve the highest possible performance. When reporting benchmarks results in research papers, it is worth considering different lexical search implementations in addition to \texttt{BM25S}.

Additionally, in order to ensure reproducibility and accessibility, our experiments are all performed on free and readily available hardware (\cref{sec:appendix}). As a result, experiments that are less memory efficient terminated with OOM errors. 

\section*{Acknowledgements}

The author thanks Andreas Madsen and Marius Mosbach for helpful discussions.

\bibliography{custom}

\appendix

\section{Appendix}
\label{sec:appendix}

\paragraph{Hardware}
To calculate the queries per second, we run our experiments using a single-threaded approach. In the interest of reproducibility, our experiments can be reproduced on Kaggle's free CPU instances\footnote{\url{https://www.kaggle.com/}}, which are equipped with a Intel Xeon CPU @ 2.20GHz and 30GB of RAM. This setup reflects consumer devices, which tend have fewer CPU cores and rarely exceed 32GB of RAM.

\paragraph{BEIR Datasets} BEIR \cite{thakur2021beir} contains the following datasets: Arguana (AG; \citealp{ArguAna_Wachsmuth2014ARC}), Climate-FEVER (CF; \citealp{CLIMATE_FEVER_Diggelmann2020CLIMATEFEVERAD}), DBpedia-Entity (DB; \citealp{DBpediaEntity_Hasibi2017}), FEVER (FV; \citealp{FEVER_Thorne18Fever}), FiQA (FQ; \citealp{FIQA_Maia2018WWW18OC}), HotpotQA (HP; \citealp{HotpotQA_Yang2018}), MS MARCO (MS; \citealp{MSMARCO_2016}), NQ (NQ; \citealp{NQ_kwiatkowski2019natural}), Quora (QR)\footnote{\url{https://quoradata.quora.com/First-Quora-Dataset-Release-Question-Pairs}}, SciDocs (SD; \citealp{SciDocs_Cohan2020SPECTERDR}), SciFact (SF; \citealp{SciFact_Wadden2020FactOF}), TREC-COVID (TC; \citealp{TREC_COVID_2020}), Touche-2020 (WT; \citealp{webis_touche_2020_bondarenko2020touche}).

\end{document}